\documentclass[doublecol]{epl2}

\usepackage{graphicx}
\usepackage{amsmath}
\usepackage{amssymb}

\title{ Spectrum of Radiation from Rough Surfaces}

\author{Zh.S. Gevorkian }
\institute { Institute of Radiophysics and Electronics,Ashtarak-2,0203,Armenia.}


\abstract{Radiation from a charged particle travelling parallel to a
rough surface has been considered. Spectral-angular
intensity is calculated in the weak scattering regime. It is shown
that the main contribution to the radiation intensity is determined by the
multiple scattering of polaritons induced by a charge on the
surface. Multiple scattering effects lead to strong frequency
dependence of radiation intensity. Possible applications in beam
and surface diagnostics are discussed.}

\begin{document}

\maketitle

\section{Introduction}

 Light scattering from rough surfaces attracted much interest \cite{ODM87}. In particular this interest is caused by the determination of microtopographic
properties of rough metallic surfaces from the light scattering
measurements \cite{VG09}. The enhancement of intensity of scattered
light on a rough surface is due to a resonant excitation
of surface polaritons induced by incident light. Surface polaritons
are multiply scattered on the roughness resulting in their
diffusion and localisation\cite{ASB85} that lead to peculiarities
in the light scattering from rough surfaces.

Polaritons can be induced not only by an incident light but also by a charged
particle. It is interesting to
reveal the manifestation of polaritons multiple scattering on the charged particle radiation from rough surfaces.
 Origination of this radiation
is due to the scattering of polaritons induced by
the charged particle on the inhomogeneites of dielectric
constant associated with the roughness of the surface.
Earlier in this geometry main attention was paid to the
periodical grating case when Smith-Purcell radiation(SPR)
\cite{SP53} is originated. Recently we have considered \cite{Gev10}
the radiation from an uncorrelated rough surface.

In the present paper we consider the influence of multiple scattering
effects, including the localisation of polaritons, on the radiation of
a charged particle travelling over a correlated rough surface.We will see that they lead to strong frequency dependence of intensity. Strong
frequency dependence allows to separate the diffusional mechanism of
radiation from other radiation mechanisms.

\section {Formulation of the Problem}
A charged particle moves uniformly in the vacuum at the
distance $d$ from the plane $z=0$  separating vacuum and the
isotropic medium  distorted by roughness.
We are interested in radiation field far away
from the charge and interface. Maxwell equation for the electric
field has the form
\begin{equation}
\nabla^2\vec E(\vec r,\omega)-\rm{grad}div\vec E(\vec r,\omega
)+\frac{\omega^2}{c^2}\varepsilon(\vec r,\omega)\vec E(\vec
r,\omega)=\vec j(\vec r,\omega)
 \label{Max1}
\end{equation}
where $\vec j(\vec r,\omega)=-\frac{4\pi ie\omega\vec
v}{vc^2}\delta(z-d)\delta(y)e^{i\omega x/v}$ is the current density associated with the charge. Here $\vec v$ is the velocity of the particle  moving on $0x$ direction and $\varepsilon(\vec r,\omega)$ is the inhomogeneous
dielectric permittivity of the system. For a rough surface it can
be chosen as $\varepsilon(\vec
r,\omega)=\varepsilon_0(z,\omega)+\varepsilon_r(\vec r,\omega)$,
where $\varepsilon_0(z,\omega)=\Theta(z)+
\Theta(-z)\varepsilon(\omega)$ descibes the flat interface
vacuum-metal and $\varepsilon_r(\vec
r,\omega)=[\varepsilon(\omega)-1]\delta(z)h(x,y)$ is the
contribution of small roughness. $h(x,y)$ is the amplitude
of surface roughness.
To separate the radiation field one should decompose the electric field as $\vec E= \vec E_0+\vec E_r$, analogous to the decomposition of dielectric constant. $\vec E_0$ and $\vec E_r$ are the background  field created by the charge and the radiation field,respectively. They obey the following equations
\begin{eqnarray}
&&\nabla^2\vec E_0(\vec r,\omega)-\rm{grad}div\vec E_0(\vec r,\omega)+
\nonumber \\
&&+\frac{\omega^2}{c^2}\varepsilon_0(z,\omega)\vec E_0(\vec
r,\omega)=\vec j(\vec r,\omega)\label{maxb1}\\
&&\nabla^2\vec E_r(\vec r,\omega)-\rm{grad}div\vec E_r(\vec r,\omega
)+\frac{\omega^2}{c^2}\varepsilon(z,\omega)\vec E_r(\vec
r,\omega)+\nonumber\\
 &&+\frac{\omega^2}{c^2}\varepsilon_r(\vec r,\omega)\vec E_r(\vec
r,\omega)=-\frac{\omega^2}{c^2}\varepsilon_r(\vec r ,\omega)\vec
E_0(\vec r,\omega)
 \label{maxbacrad}
\end{eqnarray}
Radiation intensity at the
frequencies $\omega$, $\omega+d\omega$ and at the angles
$\Omega$, $\Omega+d\Omega$ is determined as  $dI(\omega,\vec n)=\frac{c}{2}|\vec E_r(\vec R)|^2R^2d\Omega
d\omega$, where $\vec n$ is unit vector on the direction of observation
point $\vec R$, $\Omega$ is the corresponding solid angle, see
\cite{Gev10}.  Radiation intensity  should be averaged
over the realizations of random roughness $h(x,y)$. For this
reason it is convenient to introduce the Green's functions of
Eqs.(\ref{maxb1},\ref{maxbacrad}).
One can represent the averaged radiation intensity tensor
$<I_{ij}(\vec R)>=<E_{ri}(\vec
 R)E^*_{rj}(\vec R)>$ in the form
\begin{eqnarray}
<I_{ij}(\vec R)>=\frac{\omega^4}{c^4}\int d\vec r d\vec r^{\prime}
 <G_{i\mu}(\vec R,\vec r)\varepsilon _r(\vec
r)\nonumber\\
G_{\nu j}^{*}(\vec r^{\prime},\vec R)\varepsilon
_r(\vec r^{\prime})> E_{0\mu}(\vec r)E_{0\nu}^{*}(\vec r^{\prime})
\label{radten}
\end{eqnarray}
where the background electric field $E_{0\mu}(\vec r)=
\int d\vec r_1 G^0_{\mu\lambda}(\vec r,\vec
r_1)j_{\lambda}(\vec r_1)$ is
expressed through the bare Green's function.
Here $<...>$ means an averaging over the surface random profile
$h(x, y)$. Note that in the original Smith-Purcell experiment
\cite{SP53} as well as in subsequent works on SPR mainly a periodical
grating in one direction is used. In this case $h( x,y)\equiv h(
x)$ is a some periodical function of one coordinate. In the
present paper  we consider correlated random grating case.
We suppose that $h$ is  a gaussian stochastic process
characterized by two parameters $<h(\vec\rho)>=0 ,
 <h(\vec\rho_1)h(\vec\rho_2)>=\delta^2W_0(|\vec \rho_1-\vec\rho_2|)$,
where $\vec \rho$ ia a two dimensional vector in the $xy$ plane,
$\delta^2=<h^2(\vec\rho)>$ is the average deviation of surface
from the plane $z=0$. Correlation function $W_0$ is characterized by
a correlation length $\sigma$ at which it is essentially
decreased.
Maxwell equations for electric fields
Eq.(\ref{maxb1},\ref{maxbacrad}) and Green's functions
should be amended by the boundary
conditions. As usual, it is required that tangential components of
electric field be continues across the plane $z=0$. The exact
field, of course, will satisfy the boundary conditions across the
surface $z=h(x,y)$ rather than the plane. However this
approximation seems reasonable for small roughness $\lambda\gg
\delta$ and is widely used in  literature. The Green's function
$G_{\mu\nu}(\vec r,\vec r^{\prime},\omega)$, when considered a
function of $z$ for fixed $z^{\prime}$ satisfies the same boundary
condition as the $\mu th$ Cartesian component of electric field.

\section{ Green's Functions} The equation for bare Green's function
 with the correct boundary conditions for arbitrary
$\varepsilon(\omega)$ was solved in \cite{MM75}. To obtain
radiation intensity in vacuum we will need Green's functions in
the half space $z>0$. In order to simplify the problem we will
consider the case when isotropic medium is a metal with very large
negative dielectric constant $|\varepsilon(\omega)|\gg 1$. In
the limit $|\varepsilon|\to \infty$ the following components
survive \cite{MM75}

\begin{eqnarray}
G_{zz}^0(\vec p|0,z)=G_{zz}^0(\vec
p|z,0)=\frac{ip^2}{k^2}\frac{\varepsilon(\omega)e^{iqz}}{k_1-\varepsilon(\omega)q}\nonumber \\
 G_{xz}^0(\vec p|0,z)=-G_{zx}^0(\vec
p|z,0)=-\frac{ip_x}{k^2}\frac{\varepsilon(\omega)qe^{iqz}}{k_1-\varepsilon(\omega)q}
\label{grfun2}
\end{eqnarray}
where $G_{ij}^0(\vec p|z,z^{\prime})$ is the two-dimensional
Fourier transform of $G_{ij}^0(\vec r,\vec r^{\prime})$ and $z>0$.
Here $\vec p$ and $\vec \rho$ are two-dimensional vectors with
Cartesian components $p_x,p_y,0$ and $x,y,0$. Also $k=\omega/c$,
$k_1$ and $q$ are determined as follows:
\begin{eqnarray}
q=\left\{ \sqrt{k^2-p^2},\quad k^2>p^2 \atop i\sqrt{p^2-k^2},\quad
k^2<p^2 \right. \\
k_1=-(\varepsilon(\omega)k^2-p^2)^{1/2}
 \label{wavenum}
\end{eqnarray}
A branch cut for the square root in Eq.(\ref{wavenum}) along the negative real axis is assumed \cite{MM75}. Other components of
Green's function are small for large $|\varepsilon|$.
 To determine radiation intensity we will need asymptotics of
Green's functions at large distances. Using
Eq.(\ref{grfun2}) and making a Fourier transform, one finds \cite{Gev10}
\begin{eqnarray}
&&G_{zz}^0(\vec R,\vec
\rho,0)\approx\frac{1}{2\pi\sqrt{2}R}\left[n_z\sqrt{n_{\rho}}\cos\left(
k(R-\vec n_{\rho}\vec \rho)-\frac{\pi}{4}\right) +\right.\nonumber\\
&&\left.+\frac{n_z}{\sqrt{n_{\rho}}}\cos\left(k(R-\vec
n_{\rho}\vec
\rho)+\frac{\pi}{4}\right)\right]+\nonumber\\
&&+\frac{i}{2\pi\sqrt{2}R}\left[\sqrt{n_{\rho}}\cos\left(k(R-\vec
n_{\rho}\vec\rho)+\frac{\pi}{4}\right)-\right.\nonumber\\
&&\left.-\frac{1}{\sqrt{n_{\rho}}}\cos\left(
k(R-\vec n_{\rho}\vec \rho)-\frac{\pi}{4}\right)\right];\nonumber\\
&&G_{xz}^0(\vec R,\vec \rho,0)=-G_{zx}^0(\vec \rho,0,\vec
R)\approx\nonumber\\
&&\approx\frac{1}{2\pi\sqrt{2}R}\left[n_x\sqrt{n_{\rho}}\sin\left(
k(R-\vec n_{\rho}\vec \rho)+\frac{\pi}{4}\right) +\right.\nonumber\\
&&\left.+\frac{n_x}{\sqrt{n_{\rho}}}\sin\left(k(R-\vec
n_{\rho}\vec
\rho)-\frac{\pi}{4}\right)\right]+\nonumber\\
&&+\frac{i}{2\pi\sqrt{2}R}\left[\sqrt{n_{\rho}}n_xn_z\sin\left(k(R-\vec
n_{\rho}\vec\rho)-\frac{\pi}{4}\right)-\right.\nonumber\\
&&\left.-\frac{n_zn_x}{\sqrt{n_{\rho}}}\sin\left( k(R-\vec
n_{\rho}\vec \rho)+\frac{\pi}{4}\right)\right]
 \label{asym}
\end{eqnarray}
where $\vec n$ is the unit vector on the direction of the
observation point $\vec R=\vec n R$, $n_{x,z}$ and $n_{\rho}$ are
it's corresponding components. Eqs.(\ref{asym}) are correct
provided that $kR\gg 1$, $R_{\rho}\gg \rho$ and we use approximate
equation $|\vec R-\vec r|\approx R-\vec n\vec r$.

\section{ Radiation Intensity} Spectral-angular radiation intensity
Eq.(\ref{radten}) can be represented as a sum of three contributions,
$I(\vec R,\omega)=I^0(\vec R,\omega)+I^D(\vec R,\omega)+
I^C(\vec R,\omega)$, where
$I^0$ , $I^D$ and $I^C$ are single scattering , diffusive  and maximally
crossed diagram contributions,
respectively \cite{Gev98,Gev00,Gev06}. First consider the single scattering
contribution to the radiation intensity. Substituting the Green's
functions in  Eq.(\ref{radten}) by the bare ones, we obtain
\begin{eqnarray}
I_{ij}^0(\vec R)=\int d\vec \rho d\vec
\rho^{\prime}G^0_{iz}(\vec R,\vec
\rho,0)G^{*0}_{zj}(\vec \rho^{\prime},0,\vec R)\nonumber\\
 W(|\vec\rho-\vec
\rho^{\prime}|)E_{0z}(\vec \rho,0)E^*_{0z}(\vec \rho^{\prime},0)
\label{sincon}
\end{eqnarray}
where $(ij)\equiv(xz)$ and
$W(\rho)\equiv (\varepsilon-1)^2k^4\delta^{2}W_0(\rho)$.

The background electric field in the limit $|\varepsilon|\to \infty$
can be found from Eq. (\ref{grfun2}) and the form of the current density
\begin{equation}
E_{0z}(\vec
\rho,0)=-\frac{4ee^{ik_0x}}{v}\frac{dk_0}{\gamma\sqrt{y^2+d^2}}K_1(\frac{k_0\sqrt{y^2+d^2}}{\gamma})
\label{bacfil2}
\end{equation}
where $k_0=\omega/v$, $\gamma=(1-v^2/c^2)^{-1/2}$ is the Lorentz
factor of the particle and $K_1$ is the first order Macdonald
function. As  follows from Eq.(\ref{bacfil2}) the background
electric field and correspondingly radiation intensity is
exponentially small when $k_0d/\gamma\gg 1$.
An essential intensity exists for $k_0d/\gamma\ll
1$. Far away from the system at the observation point one can use
asymptotic expressions for Green's functions Eq.(\ref{asym}).
Substituting Eqs.(\ref{asym})  into
Eq.(\ref{sincon}), for the spectral-angular radiation intensity
$I(\omega,\Omega)=cR^2I_{ii}(\vec R)/2$, one obtains
\begin{eqnarray}
&&I^0(\omega,\Omega)=\frac{c}{32\pi^2}
\frac{(1+n_{\rho}^2)(1+n_z^2)(1-n_x^2)}{n_{\rho}}\times
\nonumber\\
&&\times\int dx dx^{\prime}dy dy^{\prime}\cos[k\vec n_{\rho}
(\vec \rho-\vec\rho^{\prime})]
W(|(\vec \rho-\vec\rho^{\prime})|)\times\nonumber \\
&&\times E_{0z}(\vec \rho,0)E^*_{0z}(\vec \rho^{\prime},0)
\label{sicon1}
\end{eqnarray}

Substituting background electric field Eq.(\ref{bacfil2})
into Eq.(\ref{sicon1}) and taking a gaussian form for correlation
function $W_0(\rho)=e^{-\rho^2/\sigma^2}$ after calculation of
integrals, for single scattering contribution, one has
\begin{eqnarray}
&&I^0(\omega,\Omega)=\frac{e^2}{c\beta^2}\frac{(\varepsilon-1)^2k^4
\delta^2\sigma}
{4\pi^{3/2}}\times\\
&&\times\frac{L_x(1-n_x^2)(1+n_z^2)(1+n_\rho^2)}{
n_{\rho}d}\times\nonumber\\
&&\times\left[e^{-\frac{\omega^2\sigma^2(n_x+1/\beta)^2}{c^2}}+
e^{-\frac{\omega^2\sigma^2(n_x-1/\beta)^2}{c^2}}\right] F(\frac{dk_0}{\gamma},d,\sigma,kn_y)\nonumber
\label{sicon2}
\end{eqnarray}
where $L_x$ is the system size in the $x$ direction,
  $\beta=v/c$ and $F$ is determined as follows:
\begin{eqnarray}
&& F(\frac{dk_0}{\gamma},d,\sigma,kn_y)=(\frac{dk_0}{\gamma})^2
  \int dydy^{\prime}e^{ikn_y(y-y^{\prime})}
\nonumber\\
&&e^{-(y-y^{\prime})^2/\sigma^2}  \frac{K_1(\frac{k_0\sqrt{y^2+d^2}}{\gamma})
  K_1(\frac{k_0\sqrt{y^{\prime2}+d^2}}{\gamma})}{\sqrt{(y^2+d^2)
  (y^{\prime2}+d^2)}}
  \label{ef}
  \end{eqnarray}
 When obtaining Eqs.(\ref{sicon2}) and (\ref{ef}) we neglect strongly
oscillating terms in the limit $kR\gg 1$.
The components of unit vector
$\vec n$ are determined through the polar $\theta$ and azimuthal
$\phi$ angles of observation direction: $n_z=\cos\theta,
n_{\rho}=\sin\theta, n_x=\sin\theta \sin\phi$. We are considering
radiation into the half-space $z>0$ (vacuum), hence
$\theta<\pi/2$.
Because the exponential factors in Eq.(\ref{sicon2})an essential
radiation is emitted provided that
\begin{equation}
\frac{\omega\sigma}{c}(n_x\pm 1/\beta) \lesssim 1
\label{spgen}
\end{equation}
Eq.(\ref{spgen}) generalizes the Smith-Purcell dispersion
relation \cite{Gev10} to correlated rough surface case. Now let us consider
the asymptotics of $F$ for "white noise" ($\sigma\to 0$)
and "periodical" ($\sigma\to \infty$) cases. For $\sigma\to 0$,
in the relativistic limit $k_0d/\gamma\ll
1$, substituting Macdonald function by its asymptotical expression,
one  finds, $F\approx \pi^{3/2}\sigma/4d$.
Using this,  for single
scattering contribution into spectral-angular radiation
intensity in the "white noise" case, one finds
\begin{equation}
I^{0}(\omega,\theta,\varphi)=\frac{ge^2}{\beta^2c}
\frac{(1+n_{\rho}^2)(1+n_z^2)(1-n_x^2)L_x}{8n_{\rho}d}
\label{whnois}
\end{equation}
where $g=k^4(\varepsilon-1)^2\delta^2\sigma^2$ is a
dimensionless parameter. Note that this result up to numerical
factors coincides with that in \cite{Gev10}. The difference is
caused by the definition of the correlation length and the
"white noise" limit $\sigma\to 0$.
In the opposite ''periodical" limit $\sigma\to \infty$ we find
from Eq.(\ref{ef})
\begin{equation}
F=d^2\left|\int_0^{\infty}\frac{dye^{ikn_y}}{y^2+d^2}\right|^2
\label{period}
\end{equation}
Particularly, in the most interesting case $n_y=0$, one has
$F=\pi^2/4$.
 From the condition $R_{\rho}\gg \rho$,
one obtains a restriction on angles $\sin\theta\gg L/R$, where $L$
is a characteristic size of the system.

\section{Surface Polariton }Averaged over the random surface profile Green's function of the surface  polarition satisfies the Dyson equation:
\begin{eqnarray}
&&G_{\mu\nu}(\vec p)=G^0_{\mu\nu}(\vec p)+G^0_{\mu m}(\vec
p)\int\frac{d\vec q}{(2\pi)^2}G^0_{mn}(\vec q)\times\nonumber \\
&&\times W(|\vec q-\vec p|)G_{n\nu}(\vec p)
\label{Dyson}
\end{eqnarray}
Remind that $G{\mu\nu}(\vec p)\equiv G_{\mu\nu}(\vec p|0^+,0^{+})$,
and $\vec p, \vec q$ are two dimensional vectors
see Eq.(\ref{grfun2}). Bare Green's functions are determined by
Eq.(\ref{grfun2}). In the weak scattering limit $g\to 0$, the
integral in Eq.(\ref{Dyson})is determined by the behavior of the
Green's function around its pole. As it follows
from Eq.(\ref{grfun2}) two-dimensional Green's functions of
surface polariton has a pole at
$p^2=k^2\varepsilon/(\varepsilon+1)$, see \cite{MM85}. The
corresponding velocity of the surface polariton is equal to
$c\sqrt{(\varepsilon+1)/\varepsilon}<c$. Remind that we consider
the case when $\varepsilon \ll -1$. Close
to the pole and for large negative $\varepsilon(\omega)$ the
solution of Dyson Eq.(\ref{Dyson}) for Green's functions
 of the surface polariton reads

\begin{eqnarray}
G_{zz}(p)\simeq
\frac{-k}{\sqrt{-\varepsilon_1(\omega)}}\frac{1}{k^2-p^2-i{\rm Im}\Sigma(\vec p)}
\nonumber \\
G_{zx}(\vec p)\simeq
\frac{ip_x}{\sqrt{-\varepsilon_1(\omega)}}\frac{1}{k^2-p^2-i{\rm Im}\Sigma(\vec p)}
\label{ave}
\end{eqnarray}
where
\begin{equation}
{\rm Im}\Sigma(\vec p)=\frac{k}{\sqrt{-\varepsilon_1}}\int \frac{d\vec q}{(2\pi)^2}
{\rm Im}
G_{zz}^0(q)W(|\vec p-\vec q|)
\label{soldys}
\end{equation}
Both the real part of the
integral in Eq.(\ref{soldys}) and the integral with
$G_{zx}(\vec p$ ) lead to renormalization of the parameters and
do not affect the pole structure Eq.(\ref{ave}). Substituting
$W(p)=ge^{-p^2\sigma^2/4}$ into Eq.(\ref{soldys}) and
calculating the integral one has ${\rm Im}\Sigma(p)=\frac{k^2g}{4|\varepsilon_1|}e^{-\frac{k^2\sigma^2+
p^2\sigma^2}{4}}I_0\left(\frac{kp\sigma^2}{2}\right)$,
where $I_0$ is the Bessel function. Surface polariton mean
free path on the rough surface is determined as $l=k/Im\Sigma(k)$.
Note that our consideration is correct provided that
$\lambda/l\ll 1$ \cite{AGD63}. One can find asymptotics of $l$ for
short wavelength and long wavelength regions
\begin{eqnarray}
l=\left\{ \frac{4|\varepsilon_1|}{kg},\quad \sigma\ll \lambda \atop
\frac{4|\varepsilon_1|}{kg}\frac{k\sigma}{\sqrt{2}},\quad \sigma\gg \lambda \right. \label{el}
\end{eqnarray}
As it seen from Eq.(\ref{el}) the polariton mean free
path is smaller in the long wavelength region $\lambda\gg  \sigma$.
We will see below that radiation intensity accordingly will be
larger at the same region.

\section{Diffusive Contribution to Radiation Intensity.}

\begin{figure}
\includegraphics[width=8.4cm]{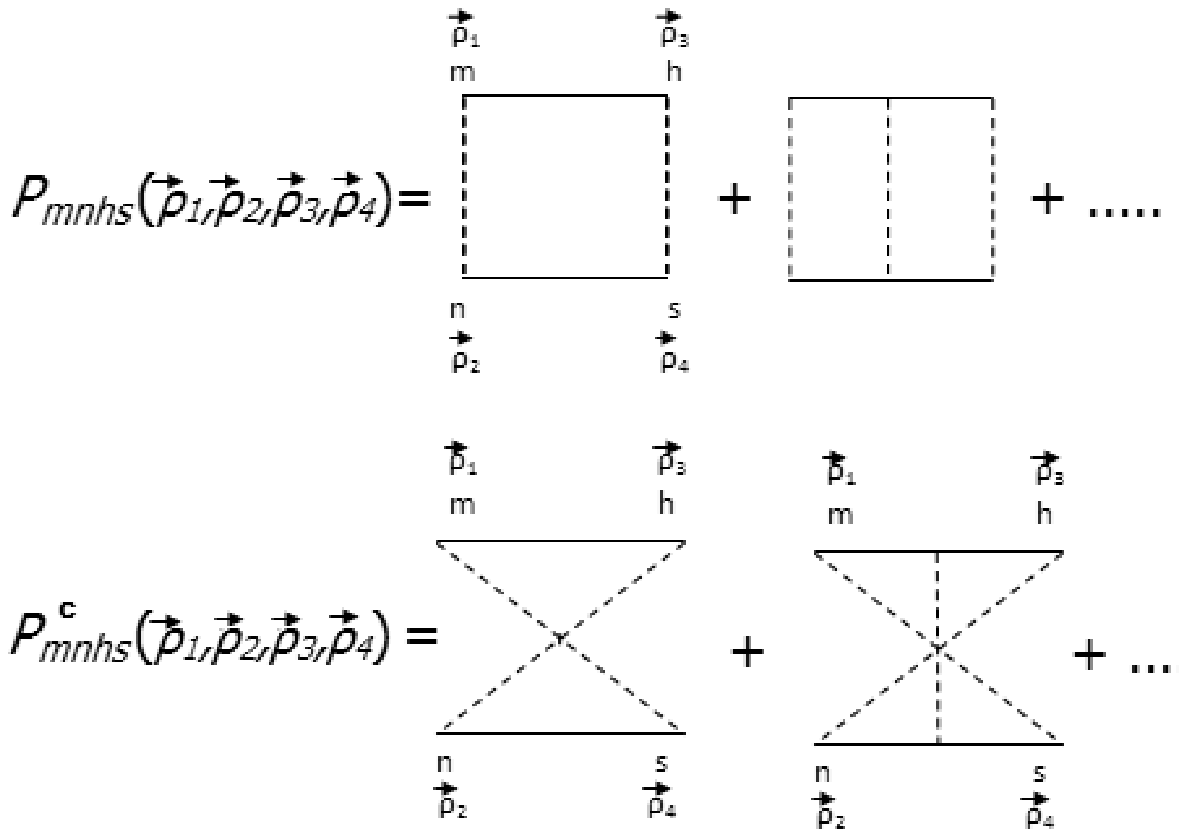}
\caption{Ladder and maximally crossed diagrams.Dashed line  is the correlation function of roughness
$W(\vec\rho_1-\vec\rho_2)$ and the solid line is the
averaged over the randomness two-dimensional Green's function of
surface polariton. }
\label{fig.2}
\end{figure}

Diffusive contribution to the radiation intensity is
determined as follows
\begin{eqnarray}
&&I^{D}(\vec R)=\int d\vec \rho d\vec \rho^{\,\prime}
d\vec \rho_1  d\vec \rho_2  d\vec \rho_3  d\vec \rho_4
G_{im}^0(\vec R,\vec \rho_1,0)
\nonumber \\
&&G_{ni}^{*0}(\vec \rho_2,0,\vec R) P_{mnhs}(\vec \rho_1,\vec \rho_2,\vec \rho_3,\vec \rho_4)
G_{hz}(\vec \rho_3,\vec \rho)\nonumber\\
&&G_{zs}^{*}(\vec \rho^{\,\prime},\vec \rho_4)
\label{difcont}
\end{eqnarray}
where $\vec R$ is the observation point, $G^0$ and $G$ are the
bare and average Green's functions respectively, see
Eqs.(\ref{asym},\ref{ave}), $\vec \rho- s$ are two
dimensional vectors on the plane $xy$, $P_{mnhs}$ is the diffusive
propagator that is determined by the sum of ladder diagrams,
see Fig.(\ref{fig.2})

It follows from Fig.(\ref{fig.2}) and the symmetry that
$P_{mnhs}(\vec \rho_1,\vec \rho_2,\vec \rho_3,\vec \rho_4)$ can be represented
in the form
\begin{eqnarray}
&&P_{mnhs}(\vec \rho_1,\vec \rho_2,\vec \rho_3,\vec \rho_4)
=W(|\vec \rho_1-\vec \rho_2|)W(|\vec \rho_3-\vec \rho_4|)\nonumber\\
&&P_{mnhs}\left(\vec \rho,\vec \rho_1-\vec \rho_2,\vec\rho_3-
\vec\rho_4\right)
\label{difprop}
\end{eqnarray}
where $\vec\rho=
\frac{1}{2}(\vec\rho_3+\vec\rho_4-\vec\rho_1-\vec\rho_2)$.
In further we will investigate only the component $P_{zzzz}$
which consists of the diffusion pole and gives the main
contribution to the radiation intensity. Using
Fig.(\ref{fig.2}) and Eq.(\ref{difprop}) and
going to the new variables, see \cite{Gev98}, we obtain the following
Bethe-Salpeter integral equation

\begin{eqnarray}
&&\int \frac{d\vec p}{(2\pi)^2}\left[1-\int\frac{d\vec q}
{(2\pi)^2}f(\vec q,\vec K)W(|\vec p-\vec q|)\right]\nonumber\\
&&P(\vec K,\vec p,\vec q^{\,\prime})=f(\vec q^{\,\prime},\vec K)
\label{frep}
\end{eqnarray}
where $G\equiv G_{zz}$,$P\equiv P_{zzzz}$ and
 $f(\vec q,\vec K)=G(\vec q+\frac{\vec K}{2})
G^{*}(\vec q-\frac{\vec K}{2})$.
Substituting Eqs.(\ref{asym},\ref{difprop}) into
Eq.(\ref{difcont}) and going to the Fourier transforms, one has
\begin{eqnarray}
&&I^D(\theta,\varphi)=\frac{c}{32\pi^2}\frac{(1+n_{\rho}^2)
(1+n_z^2)(1-n_x^2)}{n_{\rho}}
\int d\vec\rho d\vec \rho^{\,\prime}
\nonumber\\
&&\frac{d\vec p_1d\vec p_2d\vec p_3}{(2\pi)^6}W(|\vec \rho-\vec \rho^{\,\prime}|)
E_{0z}(0,\vec\rho)E_{0z}^{*}(0,\vec\rho^{\,\prime})P( 0,\vec p_1,\vec p_2)
\nonumber\\
&&|G(p_3)|^2e^{i\vec p_3(\vec\rho-\vec\rho^{\,\prime})}W(|\vec p_2+\vec p_3|)
W(|k\vec n_{\rho}+\vec p_1|)
\label{difcontfur}
\end{eqnarray}
The diffusive propagator $P(\vec K,\vec p,\vec q)$ satisfies the
integral equation Eq.(\ref{frep}). In the limit $K\to 0$ one
can search its solution in the form \cite{Gev98}
\begin{equation}
P(\vec K\to0,\vec p,\vec q)=A(K)\frac{ImG(\vec p)ImG(\vec q)}{Im\Sigma(\vec q)}
\label{Kto0}
\end{equation}
where unknown function $A(K)$ should be found from Eq.(\ref{frep}).
Substituting Eq.(\ref{Kto0}) into Eq.(\ref{frep}) and expanding
$f(\vec q,\vec K)$ up to $K^2$ one finds $A(K)$ in the form
$ A(K)=32/3K^2l^2$.
When obtaining $A(K)$ we calculate integrals in the pole
approximation that give main contribution in the weak
scattering limit ${\rm Im}\Sigma\to 0$.
Substituting Eq.(\ref{Kto0}) into Eq.(\ref{difcontfur}) and
consequently integrating with help of the Word identity
Eq.(\ref{soldys}) we finally obain for the diffusive contribution
\begin{eqnarray}
&&I^D(\omega,\vec n)=\frac{4e^2}{3\pi^2c\beta^2}\frac{(1+n_{\rho}^2)
(1+n_z^2)(1-n_x^2)}{n_{\rho}}\frac{L^2L_x}{l^2}\nonumber\\
&&\frac{{\rm Im}\Sigma(kn_{\rho})}{{\rm Im}\Sigma(k)}F_{1}(\lambda,\sigma,d)
\label{findif}
\end{eqnarray}
 where $L$ is the characteristic size of the system, $L_x$ is
 the system size in the $x$ direction and
 \begin{eqnarray}
 F_1=\left(\frac{dk_0}{\gamma}\right)^2\int dx dy dy^{\prime
 }\cos(k_0x)W(x^2+(y-y^{\prime})^2)\nonumber\\
J_0(k\sqrt{x^2+(y-y^{\prime})^2)}) \frac{K_1(\frac{k_0\sqrt{y^2+d^2}}{\gamma})
  K_1(\frac{k_0\sqrt{y^{\prime2}+d^2}}{\gamma})}{\sqrt{(y^2+d^2)
  (y^{\prime2}+d^2)}}
  \label{ef1}
  \end{eqnarray}
  Divergence of diffusive intensity Eq.(\ref{difcontfur}) is
caused by the infinite system size , see also \cite{Gev98}. If one
takes into account the finite sizes the minimal momentum in the
system become equal to $K_{min}\sim 1/L$. We take into account this
aspect when obtaining Eq.(\ref{findif}) from
 Eqs.(\ref{difcontfur},\ref{Kto0}).

Comparing single scattering Eq.(\ref{sicon2}) and diffusive
Eq.(\ref{findif}) contributions, one has $I^D/I^0\sim L^2/l^2\gg
1$. Hence diffusion of surface polaritons is the main
mechanism of radiation.

 Note that $F_1$ in Eqs.(\ref{findif},\ref{ef1})does not depend on angle. It is just a number.
 First we analyze diffusive radiation intensity Eq.(\ref{findif})
 in the short wavelength region $k\sigma\gg1$. Consider the
 ratio ${\rm Im}\Sigma(kn_\rho)/{\rm Im}\Sigma(k)\approx \frac{1}
{\sqrt{n_\rho}}exp\left[-\frac{k^2\sigma^2}{4}
(1-n_\rho^2)\right]$.
Because of the
exponential function  essential intensity is emitted on
directions $n_\rho\approx 1$, that is parallel to the metal
surface. For long wavelengths $k\sigma\ll 1$, on the contrary,
maximum is achieved in the direction perpendicular to the surface .
Now consider long wavelength region
$k\sigma\ll 1$. In this case one can  substitute
$W_0(x^2+(y-y^{\prime})^2)\to \pi\sigma^2
\delta(x)\delta(y-y^{\prime})$ in  Eq.(\ref{ef1}).
After this simplification, calculating the integrals in
 Eq.(\ref{ef1}), one finds from Eq.(\ref{findif})
 \begin{equation}
 I^D(\omega,\vec n)=\frac{2e^2}{3c\beta^2}\frac{g(1+n_{\rho}^2)
(1+n_z^2)(1-n_x^2)}{n_{\rho}}\frac{L^2}{l^2}\frac{L_x}{d}
\label{whnoisD}
\end{equation}
Note that we have missed the dimensionless constant $g$ in
Eq.(34) of \cite{Gev10} when considering "white noise" case. Besides that these two expressions differ from each other by a
numerical factor. Both these expressions are correct
with accuracy up to a numerical factor because the diffusive
propagator $P(K\to 0,p,q)$ can be found only with such
accuracy.

\section{Maximally crossed diagrams contribution}

 Maximally crossed
diagrams, (see Fig.1) contribution to the radiation intensity  reads
\begin{eqnarray}
I^{C}(\vec R)=\int d\vec \rho d\vec \rho^{\,\prime}
d\vec \rho_1  d\vec \rho_2  d\vec \rho_3  d\vec \rho_4
G_{im}^0(\vec R,\vec \rho_1,0)\\\nonumber
G_{ni}^{*0}(\vec \rho_2,0,\vec R)P_{mnhs}^C(\vec \rho_1,\vec \rho_2,\vec \rho_3,\vec \rho_4)
G_{hz}(\vec \rho_3,\vec \rho)
\\ \nonumber
G_{zs}^{*}(\vec \rho^{\,\prime},\vec \rho_4)W(|(\vec \rho-\vec\rho^{\,\prime})|)
E_{0z}(\vec \rho,0)E^*_{0z}(\vec \rho^{\,\prime},0)
\label{cross}
\end{eqnarray}
Due to the time reversal symmetry propagator $P^C_{mnhs}$
 is related to the diffusive
propagator as $P^C_{mnhs}(\vec \rho_1,\vec \rho_2,\vec \rho_3,
\vec \rho_4)\equiv P_{mshn}(\vec \rho_1,\vec \rho_4,\vec \rho_3,
\vec \rho_2)$, see for example,\cite{LR85}. Calculating analogously
to the diffusive contribution case,  one has
\begin{eqnarray}
I^C(\theta,\omega)=\frac{c}{3\pi^2}\frac{(1+n_{\rho}^2)
(1+n_z^2)(1-n_x^2)}{n_{\rho}}
\int d\vec\rho d\vec \rho^{\,\prime}\\\nonumber
\frac{d\vec p_1d\vec p_2d\vec p_3}{(2\pi)^6}W(|\vec \rho-\vec \rho^{\,\prime}|)
E_{0z}(0,\vec\rho)E_{0z}^{*}(0,\vec\rho^{\,\prime})\\ \nonumber
\int\frac{d\vec K}{(2\pi)^2}\frac{{\rm Im}\Sigma(k)}{K^2
\left[(k^2-(\vec K-k\vec n_{\rho})^2)^2+{\rm Im}\Sigma^2(k)\right]}
\label{cr}
\end{eqnarray}
It follows from Eq.(\ref{cr}) that integral over $K$
logarithmically diverges at small $K$. This divergence is manifestation
of localisation effects in radiation. It is analogous to the same effects
in disordered electronic systems, see for example,\cite{LR85}.
In the weak scattering regime ${\rm Im}\Sigma\to 0$, maximal
value of $I^C$ is achieved at directions parallel
to the surface for which $n_{\rho}$ is close to unity.
Cutting integral on $K$ on the bottom limit at $1/L$ and on the upper
limit at $1/l$, we finally find from Eq.(\ref{cr})
\begin{equation}
I^C(\theta,\omega)=\frac{2ge^2}{3\pi\beta^2c}\frac{(1+n_{\rho}^2)
(1+n_z^2)(1-n_x^2)}{n_{\rho}d}\frac{1}{kl}\ln\frac{L}{l}
\label{fincr}
\end{equation}
It follows from Eq.(\ref{cr}) that the peak of angular distribution of $I^C$
around the direction $n_{\rho}=1$ has width of order $\sqrt{\lambda/l}$.
Remind that the analogous light backscattering peak from a disordered
medium in three dimensions has a peak with width $\lambda/l$,see
for example, \cite{thro99}.
Remind that our consideration is correct in the weak
scattering regime $kl\gg 1$. Therefore maximally crossed
diagrams contribution to the radiation intensity is small.
However in the light localisation regime $kl\sim 1$ \cite{ASB85}
it becomes important because leads to strong frequency dependence
of radiation intensity (see below).
 One can notice the different dependences of single scattering
 Eq.(\ref{sicon2}) and diffusive contributions Eqs.(\ref{findif},\ref{cr})
 to radiation intensity on system sizes. The reason of this
 difference is that in the first case radiation is formed as
 incoherent sum of intensities from independent scatterers and
 therefore is proportional to system size or the number of
 scatterers. In the diffusive radiation case interference
 terms play important role. They lead to a stronger dependence
 of the intensity on the system sizes. Therefore diffusive contribution
 to the radiation intensity can be considered as a type of coherent
 radiation \cite{Mik72}.

\section{Spectrum of Radiation }
 We have assumed that the absorption of electromagnetic
field is absent.
In this consideration weak $l\ll l_{in}$,where
$l_{in}=\varepsilon_1^2/k\varepsilon_2$ is the inelastic
mean free path of surface polariton, absorption can be taken into
account as follows \cite{And85}. When $L>(ll_{in})^{1/2}$,   $L$
in Eqs.(\ref{findif},\ref{cr}) should be substituted by $(ll_{in})^{1/2}$.
 As was mentioned above in the
short wavelength region $k\sigma\gg 1$ the radiation is
directed parallel to the surface and its intensity is
suppressed compared to the long wavelength $k\sigma\ll 1$
region because the largeness of polariton elastic mean free path. Hence
the long wavelength region is more interesting .
We will investigate frequency dependence of the radiation intensity in this region.
Making the above mentioned substitution and selecting  parts
depending on the frequency, for the spectral radiation intensity,
from Eqs.(\ref{whnois},\ref{whnoisD},\ref{fincr}), one has
\begin{eqnarray}
&&I_0(\omega)\sim g(\omega),\quad
I^D(\omega)\sim g(\omega)\frac{l_{in}(\omega)}{l(\omega)},\\ \nonumber
&&I^C(\omega)\sim g(\omega)\frac{c}{\omega l(\omega)}
\ln\frac{l_{in}(\omega)}{l(\omega)}
\label{depend}
\end{eqnarray}
In order to reveal the frequency dependence of the radiation
intensity one has to know the  frequency
dependences of $g,l$ and $l_{in}$.
They depend on  dielectric constant $\varepsilon(\omega)$
of isotropic medium which for a single metal is described by
Drude formulae $\varepsilon(\omega)=\varepsilon_1(\omega)+
i\varepsilon_2(\omega)=1-\omega_p^2/\omega(\omega+i\tau^{-1})$,
where $\omega_p$ and $\tau$ are the plasma frequency and the relaxation
time of conduction electrons,respectively. In the optical region
we always have $\omega\tau\gg 1$. Therefore, for the real and imaginary
parts of dielectric constant one has
$\varepsilon_1(\omega)\approx1-\omega_p^2/\omega^2$ and
$\varepsilon_2\approx\omega_p^2/\omega^3\tau$.
Substituting these dependencies into expressions for $g,l$
and $l_{in}$, we find $g\sim constant,\quad l\sim\omega^{-3},\quad l_{in}\sim\omega^{-2}$.
Correspondingly, using Eq.(\ref{depend}), one has
\begin{equation}
I_0\sim constant,\quad I^D(\omega)\sim \omega ,\quad
 I^C(\omega)\sim \omega^2\ln\omega
\label{findep}
\end{equation}

It follows from Eq.(\ref{findep}) that the single scattering
contribution to radiation intensity does not lead to any
frequency dependence $I^0(\omega)\sim constant$. In contrary
multiple scattering contributions lead to strong dependence
of radiation intensity on frequency. Note that strong frequency
dependences were observed  in early experiments \cite{Fab73}
on radiation from rough metallic surfaces.
Other radiation mechanisms such as, synchrotron,
transition, bremsstrahlung in the optical region do not lead
to essential frequency dependence. Only Cherenkov radiation
could lead to such dependence. However for
the particle moving over a  metallic surface in the
vacuum it does not exist.  This means that the diffusive
mechanism can be separated from the other radiation
mechanisms. It can be used for monitoring the beam position
in the accelerators . Increasing of intensity of the blue
part of spectrum would mean that beam have approached
to the walls of accelerator. Radiation of non-relativistic
electrons from rough surfaces can  be used for  diagnostic
of surface.
\section{Summary} We have considered the radiation emission when
a charged particle travels above a correlated rough metal surface. It
was shown that in the optical region  the
diffusive mechanism caused by multiple scattering of polaritons
on the roughness is the main one. Diffusive radiation is a type
of coherent radiation because interference effects play
important role in its formation.
Both long wavelength $\lambda\gg\sigma$ and short wavelength $\lambda\ll\sigma$
regions were investigated. In the long wavelength region radiation
is mainly emitted on the perpendicular to particle velocity
direction. In opposite in the short wavelength region  maximum
 of radiation is achieved on the parallel to surface directions.
A strong frequency dependence of radiation intensity is found. Its
possible application for monitoring of a beam position in
accelerators was discussed.

\end{document}